# The frog and the octopus: a conceptual model of software development

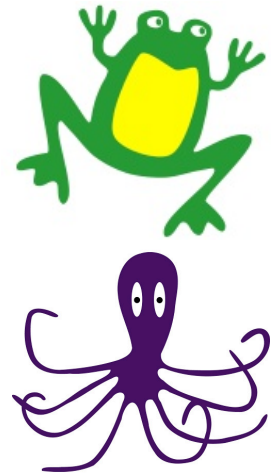

Philippe Kruchten

*University of British Columbia*
*2332 Main Mall*
*Vancouver BC V6N2T9 Canada*

email: pbk@ece.ubc.ca
phone: +1 (604) 827-5654
fax: +1 (604) 822-5949

**Abstract**
We propose a conceptual model of software development that encompasses all approaches: traditional or agile, light and heavy, for large and small development efforts. The model identifies both the common aspects in all software development, i.e., elements found in some form or another in each and every software development project (Intent, Product, People, Work, Time, Quality, Risk, Cost, Value), as well as the variable part, i.e., the main factors that cause the very wide variations we can find in the software development world (Size, Age, Criticality, Architecture stability, Business model, Governance, Rate of change, Geographic distribution). We show how the model can be used as an explanatory theory of software development, as a tool for analysis of practices, techniques, processes, as the basis for curriculum design or for software process adoption and improvement, and to support empirical research on software development methods. This model is also proposed as a way to depolarize the debate on agile methods versus the rest-of-the-world: a unified model.

**Visual abstract for JSS:**

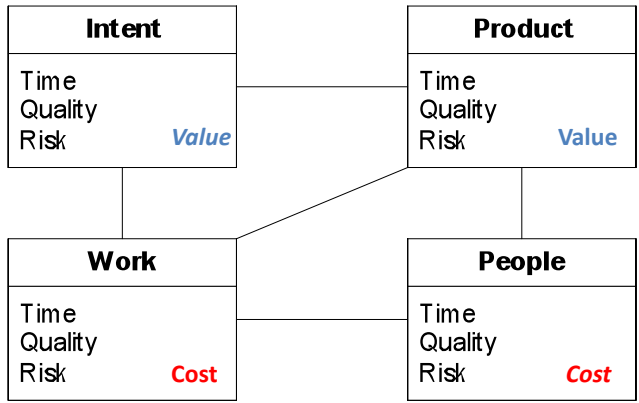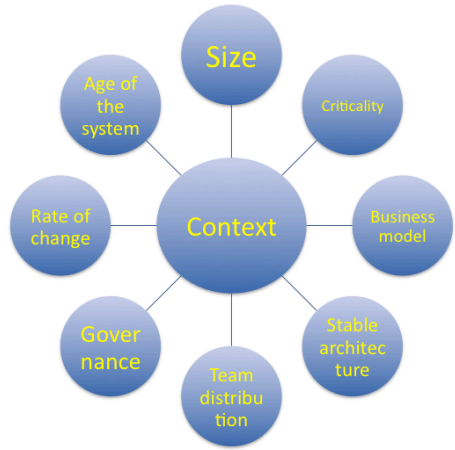

**Alternate visual abstract for JSS:**

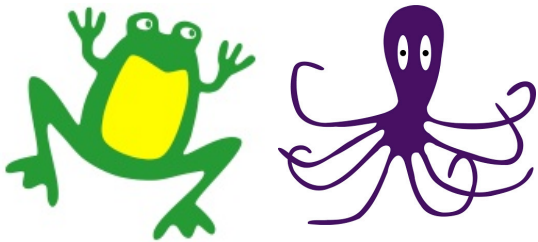

**Author's bio:**

**Philippe Kruchten** is professor of software engineering at the University of British Columbia, Vancouver, Canada. He holds an NSERC Chair in Design Engineering. He joined UBC in 2004 after some 30 years in industry, mainly with Alcatel and Rational Software, where he worked mostly with large software-intensive systems design in the domains of telecommunication, defense, aerospace and transportation. Some of his experience is embodied in the Rational Unified Process® (RUP®) whose development he directed from 1995 till 2003. His current interests are software architecture, software development processes, and software project management.

# 1. Introduction

*"The purpose of science is not to analyze or describe but to make useful models of the world."* Edward de Bono.

Software development is a complex endeavour; models are simplified abstractions of a complex reality, that allow us to reason about this complex reality. One great danger of model construction is to make the model so complex that it defeats its primary purposes, and we get entangled in debate about the model itself, and not the reality that it was originally supposed to help us cope with.

This article presents a simple model of software development. It is not revolutionary, but rather evolutionary. We have tried to absorb, evaluate (and critique) existing models, identify some of their weaknesses and limitations, and incrementally address them. To better stick in the mind of the reader, the model could be presented in the style of a *fable*, in the tradition of Aesop (circa 600 BC) or Jean de la Fontaine (1621-1695), that is, "a short story, typically with animals as characters, conveying a moral." It could start like this:

> *Once upon a time, a frog and an octopus,*
> *Met on a software project, that was deep in the bush.*
> *The frog said, "you know, all these projects are the same;*
> *Over time we fill with our work the gap that we find*
> *Between the burgeoning product, and our dreamed intent."*
> *"Oh, no" objected the octopus, "they cannot be the same;*
> *They come in all forms or shapes and sizes and colours,*
> *And we cannot use the same tools and techniques.*
> *Like in the cobbler shop, one size does not fit all."*

# 2. Motivation

The development of this conceptual model had multiple motivations. We wanted it to be:
1. Encompassing -- addressing the deficiencies of existing models, frameworks or standards, in particular with respect to the human aspects and knowledge aspects of software development.
2. Unifying -- overcoming the counterproductive polarization of the debate on software development method, especially in the last 10 years pitching agile methods *versus* rest-of-the-world, i.e., more traditional (or other) methods, while brandishing a manifesto.
3. Context-sensitive -- fighting a pervasive lack of explicit description of the development *context*: the specific circumstances, environment and constraints of a given software development project, when selecting or improving a process: lifecycle, practices, techniques and tools.
4. Coherent -- Inconsistent integration of certain concepts: risk, time, uncertainty, quality, value, cost, in many described approaches to software development.
5. Complete --There are many interesting and useful theories and models used in software engineering (Hannay et al., 2007), but they either focus on a narrow, specialized aspect of software development, or they apply to one class of software development only.

Many models have been proposed to describe, reason about, rationalize and even automate in part the software development process. Original models may have focussed mostly on the shape of the lifecycle, e.g., (Boehm, 1986; Davis et al., 1988; Larman and Basili, 2003). Since Lee Osterweil (1987) told us that "processes are software too", models of software engineering started to use the same concepts and tools as the ones we use to develop software itself: programming languages , and object-oriented models (Song and Osterweil, 1994)(Bendraou et al., 2010) for example. This has culminated with two standards: the *Software Process Engineering Metamodel* (SPEM) at (OMG, 2008), and the *Software Engineering Metamodel for Development Methodologies* (SEMDM) at (ISO/IEC, 2007).

There has been also some significant influence on our thinking about the software development process from the *Guide to the Software Engineering Body of Knowledge* (SWEBOK) (Bourque and Dupuis, 2001) and the *Guide to the Project Management Body of Knowledge* (PMBOK) (PMI, 2008) especially as the latter was adopted as IEEE software engineering standard 1490, as well as other widely used standards, such as ISO 12207 (ISO/IEC, 2008). The main difficulties with all these approaches is that they rapidly become very heavy --and this main criticism has come loudly from the agile community-- and that they imply or push organizations into rather bureaucratic (and somewhat linear) processes, in spite often the best intents (Larman et al., 2002). At their core, the paradigm is like this: software development is a kind of *factory*, where *inputs* are requirements and a sort of program is defined (the process), the *output* is the software product, and the *machine* running this 'program' is us, the humans. All we need is the proper gears and cogs, and using control systems to adjust the output as we go. This approach assumes a high level of determinism and rationality of this human "machines" and an ability to measure accurately various elements in the process. Because of their complexity, these models tend to apply better to large, complex systems, and not to the wide range of situations and contexts in which software is developed. There is little room in these models for tacit information, trust, shared mental models, or the learning process. The "thermostat" model of software development has been critiqued by many, for instance (Koskela and Howell, 2002a, 2002b).

Agile methods and agile practices are commonly introduced in a very antagonistic or even Manichaean way, as a conflict between two total opposites, the new against the old, agile versus traditional (without defining what 'traditional' really means), agility versus discipline (Boehm and Turner, 2003), agility versus waterfall, lightweight process versus heavyweight processes, agile versus plan-driven, etc. But at the same time, many mature organizations are looking at jointly using practices from the old and from the new (Vinekar et al., 2006).

Many practices, techniques or tools are often presented completely free of any context or application (Kruchten, 2007). This decontextualization, combined with a very strong propensity of our industry for confirmatory bias (i.e., only paying attention to information that confirms their preconceptions or hypotheses) lead to a lack of real debate about the value of these tools and techniques, their efficacy, and in some ways hampers research in this field (Abrahamsson et al., 2009).

Finally, there are areas of software development methods and techniques that still suffer from ambiguity, or are not fully integrated. For example there seems to be constant confusion between *value* (of a software system) and *cost* (of its development), how value and cost are estimated and how this is used to make decisions. Overall we are not lacking theories in software development, but they are often very narrow in scope, as shown by (Gregor, 2006; Hannay et al., 2007): model of uncertainties, role theory, predictive model of errors, etc.

## 3. The frog and the octopus model--a brief exposition

The model we propose is made of two parts:
1. One part, called the *frog*, presents nine universal concepts, that are found in some form or another in all software development endeavours.
2. The second part, called the *octopus*, presents eight important factors of variability across the rich spectrum of software development situations.

3. Finally we'll consider another dimension of the frog part: *tacit-explicit*, which looks at how explicit, concrete, even tangible, elements of software development are made in a given development context.

### 3.1 The frog: common elements

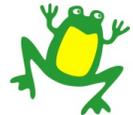

The frog says that all projects, are the same. This portion of the model shows what they all have in common. There are four core entities in our model of software development: Intent, Product, Work and People; see figure 1.

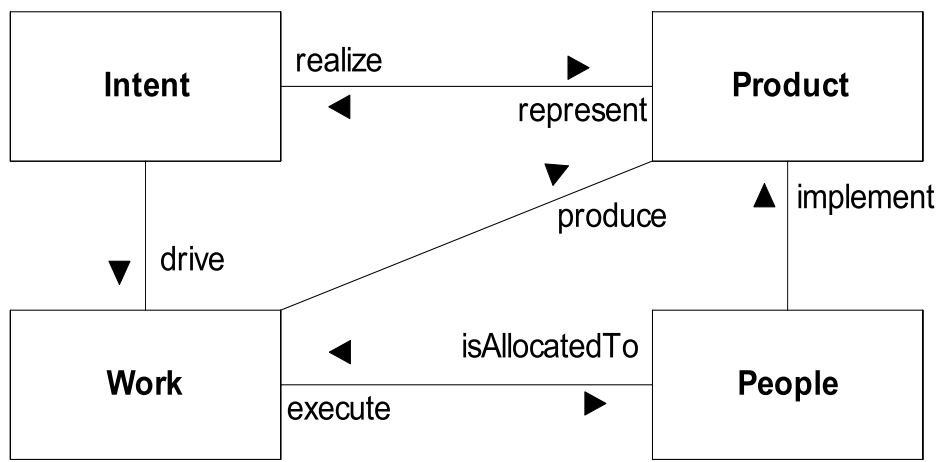

*Figure 1: Four core concepts in software development: Intent, Work, People and Product*

### 3.1.1 Intent

The concept of *Intent* denotes what the project is trying to achieve. The *Intent* defines the scope of the project, the intentions and hopes of the key stakeholders, the objectives. While we think of the intent as "the requirements" or "the specification", in practice *Intent* may take many diverse forms: a set of tests that the product must pass contributes to define *Intent*. A set of software problem reports that must be dealt with also indirectly defines *Intent*. Various constraints, implicit or explicit, internal or external to the project will also affect *Intent*. And one constant of software projects is that they are under pressure of a stream of change requests which modify the *Intent*. How much and how the Intent is made explicit will vary greatly from method to method, and over time.

### 3.1.2 Product

The concept of *Product* denotes the outcome of the project, what has been achieved. This is the actual software, accompanied with any other artifacts that are needed to make it a complete product: an installer, a set of data, the user's guide, some training material, etc. Why aren't *Intent* and *Product* more or less equivalent? Why do we need to distinguish them in our model? *Intent* precedes *Product*. *Intent* is an abstraction, a virtuality that sketches the reality that the project is set to achieve. Even when the product is "done" there may be discrepancies between the *Intent* and the *Product*; the *Intent* may have evolved in the meantime, or the *Product* could come short in some ways of the original Intent. These discrepancies --the "delta"-- between *Intent* and *Product* are the key drivers for the project; they are *the imbalance that makes the project run*. Imagine the relationship between *Intent* and *Product* as a bungee cord: the further apart they are, the more energy the project will expend to bring them closer.

### 3.1.3 Work

The concept of *Work* denotes the activities, tasks, steps that need to be accomplished in order to "turn *Intent* into *Product*". They often come defined by a process, or a method, which attempts to describe a systematic way to build a product; some elements of Work are defined "on-the-fly" in an ad hoc fashion. In many cases, a Work item produces or refines some artifact, formal or informal: a document, a model, an idea, a piece of code, a report. Some of these artifacts are only useful internally to the project, as stepping-stones, and do not appear in any form in the final product. They are not "deliverables".

### 3.1.4 People

The concept of *People* is important in modern software project management because they are the main "engine" behind *Work* elements. Software development is an intellectual activity that is very 'human-intensive'. Most of the work elements are done by human beings, and little of this work can be automated. So the availability, the competence (knowledge, skills) and the motivation of the people are keys to getting all the work done. Also most of the cost of software development is associated with people. (In figures, we will often use the word Staff and use the initial S to denote the concept of *People* and not clash with the P of product).

Each of these four core concepts has 3 attributes: Time, Quality and Risk/uncertainty.

### 3.1.5 Time
The concept of *Time* is orthogonal to our fundamental quadruple [Intent, Work, People, Product]. Often we will use the phrase *lifecycle* to denote what happens with a project over time. It is tempting to define a project linearly relative to time in 5 main steps: 1) define completely the Intent, 2) derive from the Intent all the Work that needs to be accomplished, 3) allocate work to People, and 4) People build the Product, 5) which acceptance testing will show that it matches exactly the original *Intent*. This has been tried again and again, but with very meager success in software development for a range of reasons (Kruchten, 2004; Larman and Basili, 2003). In reality, in most software development projects, we define *Intent* gradually, and it tends to evolve throughout the project under various pressures and demands for changes. We can therefore define only part of the *Work* at any point in time, and allocate it to *People*, who will therefore only build part of a *Product*. This partial product will influence back the Intent, through user feedback, or problem reports. It will also influence how people will conduct the work in the future. Other chunks of Intent are then carved out, more *Work* defined, and the *Product* will evolve until it reaches a deliverable state. Formal definitions notwithstanding, the reality of this process shows that all modern software development approaches are iterative and incremental (Larman and Basili, 2003), and they define a project as a sequence over time:

$\{$ [Intent$_1$, Work$_1$, People$_1$, Product$_1$],
   [Intent$_2$, Work$_2$, People$_2$, Product$_2$],
   …
   [Intent$_n$, Work$_n$, People$_n$, Product$_n$] $\}$

where Product$_n$ is the final 'deliverable'. Intent, Product and Work co-evolve until the product is done.

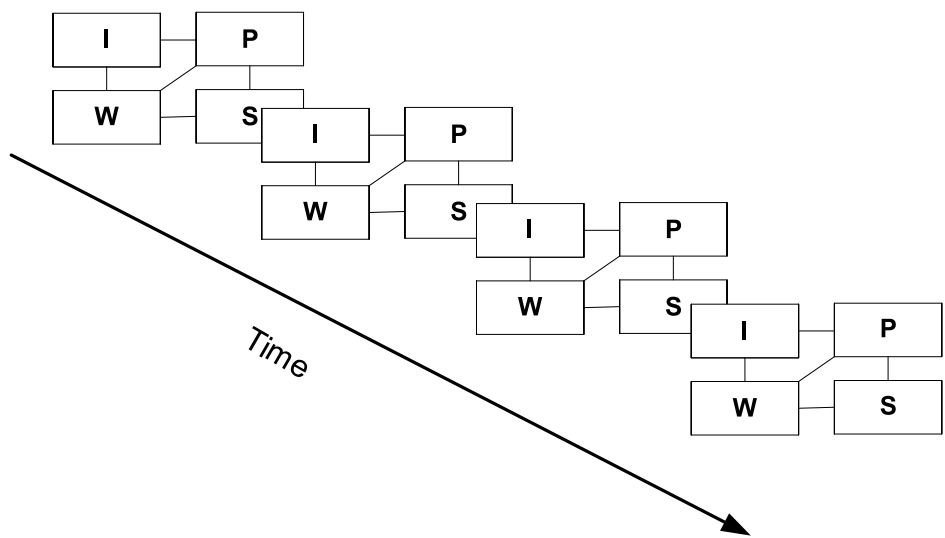

*Figure 2: Intent, Work, People (Staff) and Product evolve over Time*

### 3.1.6 Quality

The concept of *Quality* is also an orthogonal notion to our fundamental quadruple [Intent, Work, People, Product]. We can see quality as an attribute of each of them. Quality of the Intent denotes how good we are at defining and planning a Product. Quality of the Work denotes the quality of the process we use to develop software and all the intermediate artifacts, including tools, environment and management. Quality of the People denotes the knowledge, competence and diligence and dedication of the staff assigned to the project, and finally quality of the product is a measure of how close to the expectation of the stakeholders the delivered product is. These four aspects of quality may evolve over time, following the sequence we described above, and hopefully their quality increases over time (inasmuch as quality is quantifiable).

### 3.1.7 Risk and uncertainty

The concept of *Risk* denotes the uncertainty that is associated to each of the four fundamental concepts at some point in time: uncertainty in the Intent, because the domain is new, for example, uncertainty in the Work to be performed, because the process is unclear, as well as tools, environment, management, risks associated with the People and therefore uncertainties in the final Product. As with to Quality above, these uncertainties evolve over time: the risks will be mitigated, unknowns will become known, but new risks keep emerging.

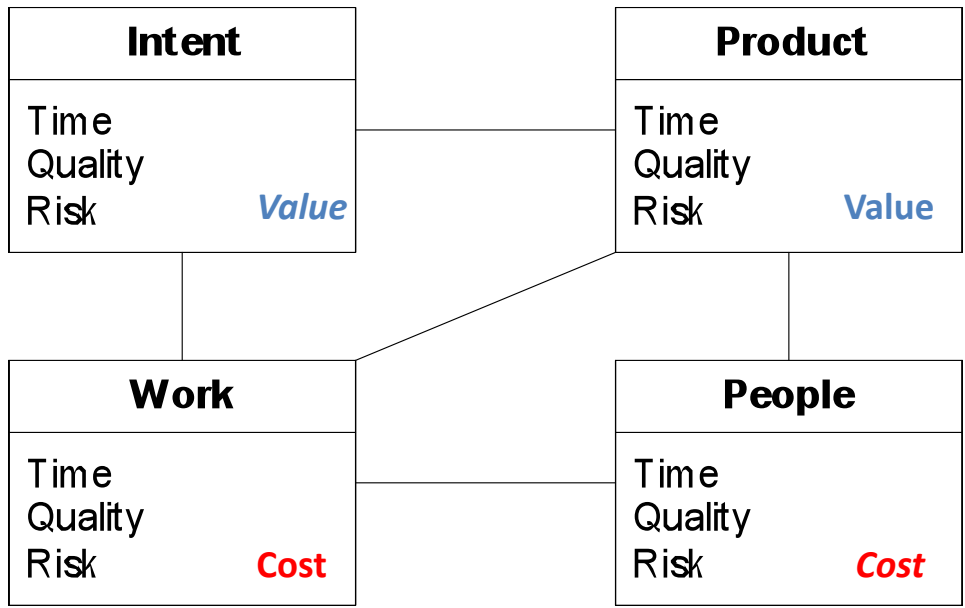

*Figure 3: The frog: Time, Quality and Risk are attributes of Intent, Work, People and Product*

### 3.1.8 Value and Cost

Finally, *Value* is associated with Intent and Product: we need to assign expected value to the Intent to guide development of the Product over time, while the *Cost* of the development is associated with the Work and the People. When we have an established Product, the value is easier to assess. As for cost, since software is essentially an intellectual, human-intensive activity, the cost is directly derived from the cost of the People associated with the project and the Work they do: what, how much, for how long. This is a characteristic of software development not shared with other engineering disciplines, such as civil engineering for example, where much of the cost is incurred outside of the development team, and the level of reuse is higher.

### 3.1.9 Project

We may now define a software development *Project* as all the *work* that *people* have to accomplish over *time* to realize in a *product* some specific *intent*, at some level of *quality*, delivering *value* to the business while incurring *costs*, and resolving many uncertainties and *risk* on the way.

The *relationships* between Intent, Product, Work and People indicated in figure 1 play an important role too, in particular in project management: who is doing this part of the work, how does the staff evolves over time, who is responsible for that subsystem, etc.

Annex A expands somewhat the frog part, showing how elements of known processes map to our 4 key entities and their 5 attributes.

## 3.2 The octopus: factors of variability

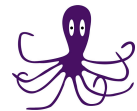

The octopus says that actually all software development projects are different. There are 2 sets of factors that make up the overall *context* of a software development project:
- factors that apply at the level of whole organization/company,
- and factors that apply at the level of the individual project.

### 3.2.1 Project-level variability factors - The octopus

Most aspects of software projects are affected by context: by size, criticality, team distribution, pre-existence of a stable architecture, governance rules, business model, that will guide which practices will actually perform best, within a certain domain and culture.

1. Size
The overall size of the system under development is by far the greatest factor, as it will drive in turn the size of the team, the number of teams, the needs for communication and coordination between teams, the impact of changes, etc. Number of person-months, or size of the code, or development budget are all possible proxies for the size.

2. Stable architecture
Is there an implicit, obvious, de facto architecture already in place at the start of the project? Most projects are not novel enough to require a lot of architectural effort. They follow commonly accepted patterns in their respective domain. Many of the key architectural decisions are done in the first few days, by choice of middleware, operating system, languages, etc. Some projects suffer from constant design volatility.

3. Business model (finance)
What is the money flow? Are you developing an internal system to support your internal processes, or a commercial product, or a bespoke system on contract for a customer, or maybe a component of a large system involving many different parties? Are you contributing to a free-libre open-source (FLOSS) project? How the various participants are ultimately compensated for their efforts will also shape the development and management process.

4. Team distribution
Linked sometimes to the size of the project, how many teams are involved and are not collocated? This increases the need for more explicit communication and coordination of decisions, as well as more stable interfaces between teams, and between the software components that they are responsible for.

5. Rate of change
Though agile methods "embrace changes" (Beck, 2000), not all domains and systems experience a very rapid pace of change in their environment: Intent volatility. How stable is your business environment and how much risk and uncertainty are you facing?

6. Age of system
Are we looking at the evolution of a large legacy system, bringing in turn many hidden assumptions regarding the architecture, or the creation of a new system with fewer constraints, a 'greenfield' project?

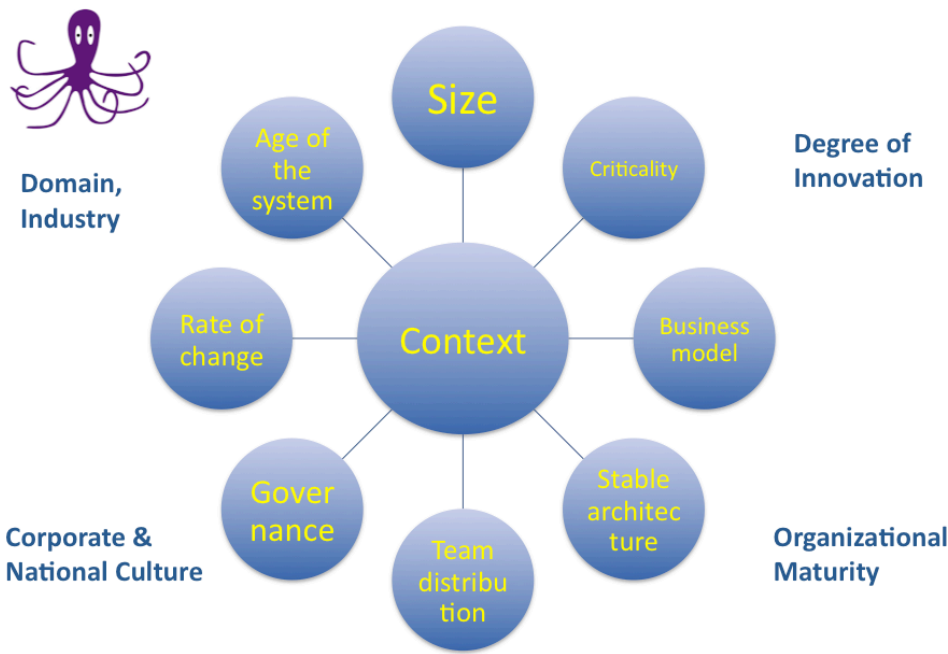

*Figure 4. Variability across software development projects (the octopus)*

7. Criticality
How many people die or are hurt if the system fails? Documentation needs increase dramatically to satisfy external agencies who will want to make sure the safety of the public is assured. For example compliance to DO178B (RTCA, 1992) for avionics bring the need for very detailed traceability between many artifacts.

8. Governance
How are projects started, terminated? Who decide what happens when things go wrong? How is success or failure defined? Who manages the software project manager? Are there external rules and regulations imposed on the product? For example the Sarbanes-Oxley act (US Government, 2002) for financial systems.

### 3.2.2 Organization-level factors

The organization-level factors (or environment conditions) heavily influence the project-level factors, which in turn exert significant influence on the processes and practices that should be used.

1. Business domain
For what domain of activity is this organization developing software? Web-based systems, aerospace embedded systems, small hand-held instrumentation?

2. Number of instances
How many instances of the software system (large or small) will be actually deployed? Are you deploying one system, a dozen, a thousand, or millions?

3. Maturity of organization
How long has that organization been developing software? How mature are the processes (and the people) relative to software development?

4. Level of innovation
How innovative is the organization? Creators or early adopters of new ideas and technologies? Or treading on very traditional grounds?

5. Culture
In which culture are the projects immersed, both national culture and corporate culture? What are the systems of values, beliefs and behaviours that will impact, support or interplay with the software development practices? Ethics is part of it.

This second set of factors often pre-determines or constrains the first one. For example, aerospace projects tend to be large, safety-critical, have complex governance rules and are not likely to be open-source; on-line e-commerce tools are not safety critical, have small teams, and a high rate of change.

### 3.3 The tacit - explicit dimension

One of the main messages of the agile movement, complemented by input from other disciplines, is that software development is a *knowledge activity* (Nonaka and Takeuchi, 1995), dealing often with rather complex, wicked problems, not easily tractable by sequential, algorithmic, procedural approaches, but often by trial-and-error, reflection and learning, iteration, the "Envision, Speculate, Explore, Adapt" loop of (Highsmith, 2004) or the Observe-Orient-Decide-Adapt loop of Boyd, cf. (Adolph, 2006). While many methods are describe primarily by artifacts (documents, code, etc) and how these artifacts are created and evolved, we do know that a large part of the knowledge handled in software development project is tacit, held primarily in the minds of the participants (Rus and Lindvall, 2002). And success in many software development projects is related to the development of *shared mental models* across the organization around the four key entities: intent, product, work and people. To be effective, the development team must develop some ways to have all four of:
1. a shared mental model of the *intent* (where is it what we want to end up)
2. a shared mental model of the current state of the *product* (and therefore the gap that remains to be filled)
3. a good understanding of the *people* in the team, their abilities, competence, knowledge and even availability,
4. leading to a clear and common understanding of the *work* that need to be accomplished and by whom.

The context (the octopus) will dictate how much of these mental models need to be supported by crutches, i.e., explicit knowledge, such as: informal or formal documents, plans, descriptions, walls of cards, communication channels, prescribed procedures and processes, tools, etc., with Size, Criticality, Governance and Geographic Distribution being in general the key drivers towards more explicit knowledge (i.e., written documents). And as shown in their SECI model (Nonaka, 1994; Nonaka and Takeuchi, 1995), there is a very complex and dynamic interplay between the tacit and the explicit elements of knowledge.

At one end of the spectrum, the *product* is the only tangible artifact, all the rest in in tacit, shared mental models, and at the very heavy bureaucratic end of the spectrum, all information is captured in some tangible artifact with no reliance whatsoever on any mental models. How explicit are the *relationships* between the four core entities (see fig. 1) is also greatly variable across the types of project.

## 4. Discussion

The core entities of the frog are not surprising and can be found in many models, though we would stress that here *people* are not just a mere little ingredient sprinkled here for good measure, the "human resource"; in software development, the People element cannot be thought as merely a pool of of substitutable, homogenous resources. Joel Jeffrey had advocated for a "more formal treatment of the human concepts" in software development, and our model is partly inspired by his (Jeffrey, 1996). His model has Wants, Knows (Facts, Concepts, Perspectives), Actions, Know-how, Process, Achievements, Personal characteristics, Significance. The frog model is a little bit simpler, and simpler to put to use in practice (see below section 5). The frog model also does not imply any specific sequences or activity, nor transformations, nor cybernetic control loop: the process stops when the product is deemed to have achieved a reasonable subset of the intent, and in this it differs somewhat from the SPEM, SEMDM, SWEBOK, PMBOK or ISO12207 paradigms. While the frog model accommodates an engineering/design model such as John Gero's *Function Behaviour Structure* (FBS) framework (Gero, 1990; Gero and Kannengiesser, 2004; Kruchten, 2005), our model can also accommodate a more emergent view of software development, such as Paul Ralph's *Sensemaking-Coevolution-Implementation Framework* (Ralph, 2010; Ralph and Wand, 2008) or the emergent metaphor of design described by (Nerur and Balijepally, 2007). Quality is not limited to the artifacts or the product as often the case when we speak about software quality, but also quality of intent, work and people. Similarly risk and uncertainty is applied equally to all 4 core entities.

This octopus model of process variability is similar to Scott Ambler's *Agile Scaling Model* (Ambler, 2009)], or the drivers Alistair Cockburn suggests for selecting a methodology (Cockburn, 2000). They also map to a certain degree with the 5 dimensions of process comparison used by (Boehm and Turner, 2003) when contrasting agile methods to more traditional ones, or those used by (Vinekar et al., 2006) when looking at reconciliation between the two camps: agile, non-agile.

The tacit-explicit dimension flows directly from work in knowledge management, and in particular the work of Nonaka and Takeuchi (Nonaka, 1994; Nonaka and Takeuchi, 1995) on the SECI model and on the notion of 'Ba' (Nonaka and Konno, 1998) (which is context under some other form). As posited by (Balijepally et al., 2007), "the social and human capital in a software development team are positively related to both the explicit and the tacit knowledge outcomes of the team," and the balance between the tacit and the explicit knowledge varies across methods, based on the context of the project.

## 5. Using the model

*"[…] a model is useful if it allows us to get use out of it."* Edward de Bono.

The frog and octopus model constitutes a sort of simple theory of software development, a theory for Analysis and Explanation -- i.e., a theory type I and II in Shirley Gregor's classification (Gregor, 2006), though we have also used it for design and action, type V. We have used the model for different purposes, which also allowed us to refine it over the years:
1. Analysis and explanation of extant processes, or their practice and techniques (I, II)
2. Building software engineering curricula (V)
3. Contribution to a body of knowledge or an ontology of software engineering (I, II, V)

4. Basis for empirical research on software engineering (V)
5. Assessing the state of a software development project.

"What would the frog say? What would the octopus say?" are now our two favourite questions when presented with a software process, a practice, a tool, or when doing a process assessment in an organization, or simply reviewing a project. Or in other words: "How do we map the elements of this process, method, tool or practice to the frog - the common elements? and how does it fit its actual context on the various dimensions of the octopus --where does it lay in the vast spectrum of software development projects? For example, we may ask:

"In your project, how do you …
- Visualize the *Intent*
- Map *People* onto *Work*
- Show *People* over *Time*
- Show responsibility of *People* onto *Product*
- Assess the *Quality* of the *Intent*
- Assess the *Risk* associated with *Intent*
- Show the *Cost* of *Work*, over *Time?*"

Or you can take a practice, a technique, an artifact, a standard, and ask:
- "Does this practice scale up to a large project?
- How would this work in a distributed team?
- What if you do this in an open-source context?
- How does it comply with a regulated environment? A safety-critical system?
- Can you skip this for a very small project?
- What is there is no architecture when you start?
- How do you control scope creep?"

A few simple examples:
- We can take the practice of a *Daily Stand Up meetings* from Scrum (Cohn, 2009). It is mostly focussed around people and the mental models they have of both *where* the project is going (Intent), and *how* it is going there (Work). It does not scale well beyond a dozen people, and is made very difficult in distributed environment, especially over multiple time-zones. Both Business model and Cultural issues may hinder it further: subcontracting, "losing face", power distance (Hofstede, 1997).
- Similarly, we can analyze a *Gantt chart*. It shows primarily how Work is spread over Time. In some forms, it can also indicate the mapping of People onto Work (task allocation). Supported by tools, it scales well, but its applicability is limited by the constant changes and the absence of clear, definite work breakdown structures (WBS) in software development.
- Going further, we can describe a *Kanban board* (Kniberg and Skarin, 2010), and then use our framework to systematically compare it with a *Gantt chart*, both in terms of functionality, how its purpose is achieved (the frog view), and in terms of applicability to various contexts (the octopus view).

The frog and the octopus model has also been used to build a new curriculum for teaching software project management (SPM); see (Kruchten, 2011a, 2011b) and software process, departing from the model of the PMBOK (PMI, 2008), and richer in scope than teaching a pure agile process like Scrum (Cohn, 2009). When teaching SPM to people who have experience in software development, as project manager or any other role, we can draw from this experience to analyze, evaluate, contrast, and discuss the role and value of various techniques. But when there is no real experience to draw from, the students see this as a mere collection of recipes to follow, with few opportunities to put them in practice. Contrasting "old" or "traditional" (meaning often: 'bad') approaches to "new" or "agile and lean" approaches proved also to be vain, as students have not experienced the "old", and as the old approaches may still be valid in some context, and good to know in general. The development of a conceptual model was therefore key in that it allowed analysis and reasoning about *any* software techniques, old or new. To our surprise, the conceptual model proved to be also as useful when teaching SPM to seasoned practitioners, in particular in explaining *why* the new, agile or lean approaches bring value, and what are the limits of their validity, and exploiting the fact the entity People is more prominent and integrated in this model than other models or frameworks.

The frog and the octopus model has been provided as an input to the SEMAT initiative (Software Engineering Method And Theory) ([www.semat.org](www.semat.org)), as well as the current attempt by IEEE and PMI to evolve the PMBOK into something more adapted to software development. As for SEMAT, our core concepts and attributes correspond to what they call "universals", i.e., entities or concepts that exist in all and every software engineering project. In undertakings like the building of any Body Of Knowledge (BOK), the impact of the model may be limited to rebalancing the weight of social and human aspects, bringing in a knowledge management dimension, and clarifying tactical issues such as value and cost, or risk.

Finally, the model can inform and provide a theoretical background to formulating research questions, helping remove some of the biases prevalent in current agile method publications, and studying software development practices (agile and non agile) in a wider range of contexts (as we proposed in (Kruchten, 2010b)). The problem is not just to demonstrate the efficacy of this or that 'new' proposed approach in the absolute, but to do so in the proper context. For a given research question, for example around Self-organizing teams (Hoda et al., 2010b; Moe et al., 2008) you can recast the question of *autonomy* (external, internal and individual) relatively to the octopus factors of Governance, or Geographic distribution, or Business model, or even Size.

## 6. Conclusion

We have presented a conceptual model of software development in two parts: the *frog*, made up of elements that are universals, common to all software projects, large or small, although they may take very different shapes, and be more or less explicit; and the *octopus*, which introduces 8 key factors that have a significant impact on the actual practices used to develop software. To sum up the model, a software development *project* is all the *work* that *people* have to accomplish over *time* to realize

15 / 26

in a *product* some specific *intent*, at some level of *quality*, delivering *value* to the business at a given *cost*, while resolving many *uncertainties and risk*. All aspects of software projects are affected by *context*: size, criticality, team distribution, pre-existence of an architecture, governance, business model, that will guide which practices will actually perform best, within a certain domain and culture.

It is our hope that this modest theory of software engineering can be used to *unify* the field of software engineering, and not simply across agile methods, but across all methods, from waterfall and heavy-weight, plan-driven, paperwork-obsessed, to nimble, lean, agile and document-less processes. This is especially interesting at a time where the community at large is revisiting some of the choices made, and discovers often that "old" practices still are serviceable arrows in their quivers.

We could add many more concepts to the model, but we feel that it is complete enough, as "… perfection is achieved not when there is nothing left to add, but when there is nothing left to take away." (Antoine de St. Exupéry, *Terre des hommes*, 1939, chap. 3) So, unless there is something in software development that does not match at all, or cannot be explained by the model, we would leave it as is. This parsimony of concept is therefore both a strength and a limitation at the same time.

Under investigation is a definition of our model as a more formal ontology, using the BBW ontological model (Bunge and Weingartner, 1974; Wand and Weber, 1995).

**Acknowledgements**
Thank you to our friends and colleagues who've provided us with valuable feedback on the model, in particular: Patrick Conroy, Rashina Hoda, Nicolas Kruchten, Eugene Nizker, Karthik Pattabiraman, and Carson Woo.

**Note to JSS reviewers:**
We did not come up with this model in one shot, and therefore bits and pieces of it have appeared here and there in other publications, courses and web sites. In particular the *frog* made a contribution to the SEMAT initiative (at www.semat.org) (Kruchten, 2010a), while the *octopus* tried to contextualize agile methods in (Kruchten, 2010b) (to be republished in the *Journal of Software Maintenance and Evolution: Research and Practice*), and it appeared briefly in (Hoda et al., 2010a), as well as in the foreword of *Agility across Time and Space* (Šmite et al., 2010, pp.v-vii). The frog+the octopus model was presented as the basis for the syllabus of a course on software project management at the CSEE&T and CEEA conferences (Kruchten, 2011a, 2011b). A short synopsis was published by the Computer Society of India.

# Annex: Expanding the frog

In this annex we detail further the four main concept in the frog part of the model. Concrete examples of elements are taken from RUP (IBM, 2007; Kruchten, 2003), XP (Beck, 2000), Scrum (Schwaber and Beedle, 2002), TDD (Beck, 2002) and various IEEE standards. Figure 6 to 9 are using UML (Rumbaugh et al., 1998). This annex also illustrate the tacit-explicit range of elements found in various approaches.

A *software project* is temporary endeavour intended to create a new software product or service, or the software part of a software-intensive system, or to evolve an existing system. It is temporary in the sense that it has a definite beginning and a definite end, in contrast with a continuous endeavour, such as running the IT operations of an organization. A software project has specific and sometimes conflicting objectives and many constraints of diverse nature, mainly technical, temporal, cultural and financial. Software project management is therefore" the art of balancing competing objectives, managing risks, and overcoming constraints to successfully delivering a product which meets the needs of both customers and users (the customer paying the bill not always being the end-user)" RUP in (IBM, 2007).

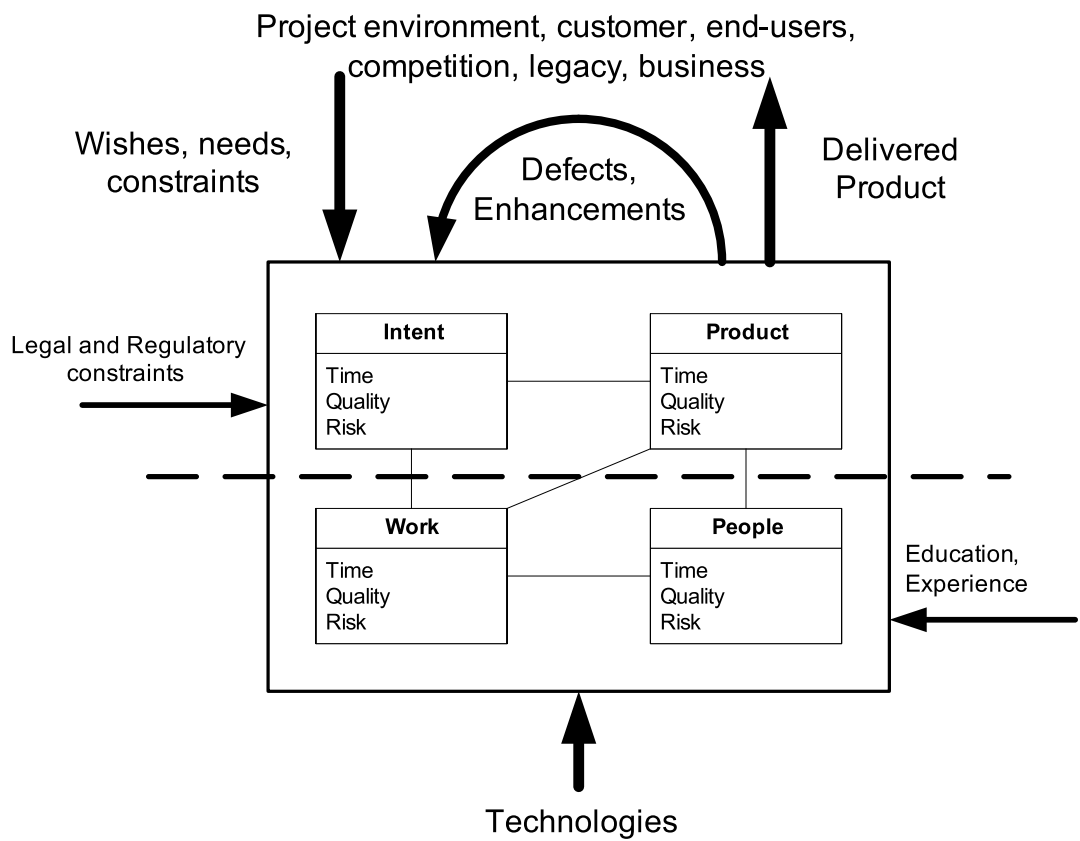

*Figure 5: The project and its context*

The software project, represented in our conceptual model with a tuple or composite object [Intent, Work, People, Product], or more precisely with a sequence of such composite objects, does not live in isolation, but it sits in a wider context, which is crucial to understand for a software project manager.

Intent and Product are mostly facing the users and customers community, intent driven mostly by them, and the product delivered to them. Constraints come from the customers, and from the business environment: the company which "owns" the project. There are also constraints coming from legal and regulatory bodies in some industries, especially in the safety-critical domains: transportation, defense, biomedical, nuclear, and in the financial domain.

People and Work are mostly influenced by the available technologies to develop and deploy the software product or service: programming languages, methods, development environment, software tools, reusable components, deployment platforms: CPU, OS, network protocols, etc.

**Intent, revisited**

The Intent of the project is an image, a description, a model of what the various parties involved want to product to be. This Intent may take several forms, depending on the type of software project and on the method or process used. We will assume that the Intent can be decomposed in a set of Intent elements, coming from various sources, and carrying different names in different methods:

- *Users needs*: a description of the needs of the user, at least the needs that we intend to satisfy
- *Vision*: a document that describes in high level terms (RUP)
- An initial product *backlog* (Scrum)
- A Software Requirement Specification, SRS (IEEE Standard 830)
- A list of user stories (XP)
- A feature list (FDD)
- A use-case model, with or without supplementary specifications for non functional requirements (RUP)
- A set of acceptance test cases (TDD)
- A list of software problem reports (or bugs, or defects)
- A prototype or mock-up
- An existing product (if we are migrating or re-engineering a legacy application, for example)

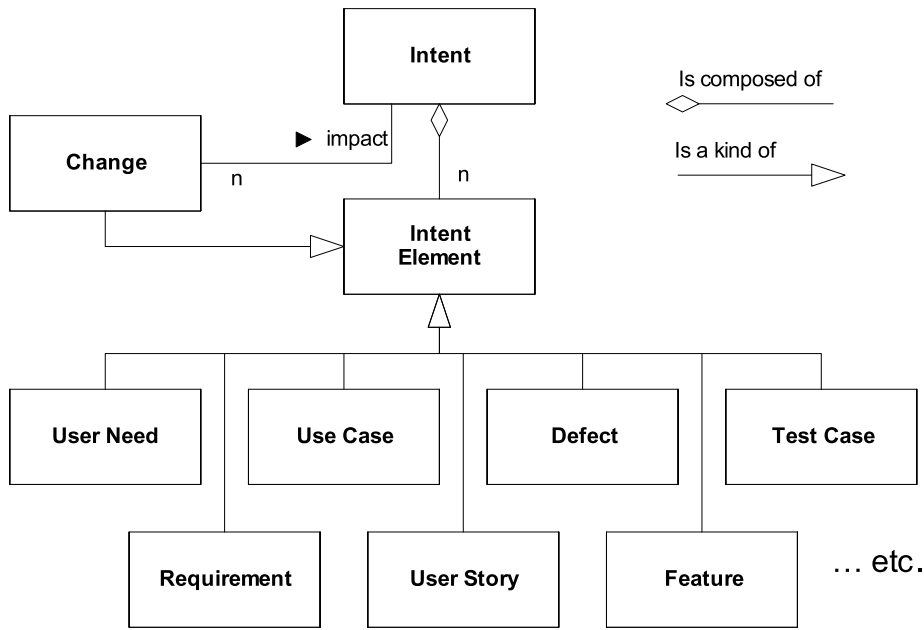

*Figure 6: Intent elements come in many different forms, and Intent constantly changes*

It's around the concept of *Intent* that we can introduce the subtle but pervasive concept of *change*, since changes occurring along time in a project are in most cases changes in Intent, which in turn will trigger changes in work and changes in the product. (There maybe also changes in people, though, not related to intent: a resignation, for example.)

**Work, revisited**
Work is the stuff that makes traditional schedules and plans with associate with project management: work items are found in network schedules, Gantt charts, in tools such as Microsoft Project®, Niku® or Primavera®. Large chunks of work make up the projects and are used in the plans, organized traditionally as Work Breakdown Schedules (WBS). Smaller work items are the items that developers put on their to-do list, scribble on their white boards or their PDAs. The bulk of process descriptions, such as RUP® (IBM, 2007) or MSF® , are dedicated to the description of work: sub-processes, activities, tasks, steps. They are the elements of focus in software process standards such as IEEE-1074 (1997) ISO 12207 (2008), or SPEM and SEMDM.
    One of the most difficult tasks of software project management is to derive from a given *Intent* the required *Work*. We still only know how to do this very approximately, and there are many work items that spring out spontaneously during the course of a project, due to unknowns, to people making errors and other various mishaps.
    The amount of Work is also a key ingredient to the estimation of effort and schedule, and therefore to the cost of the project. As we will see later, effort estimation is another big hurdle in software development, the "black art". Finally, project management attempts to monitor progress while the project is on-going by comparing actual work performed to the anticipated work.

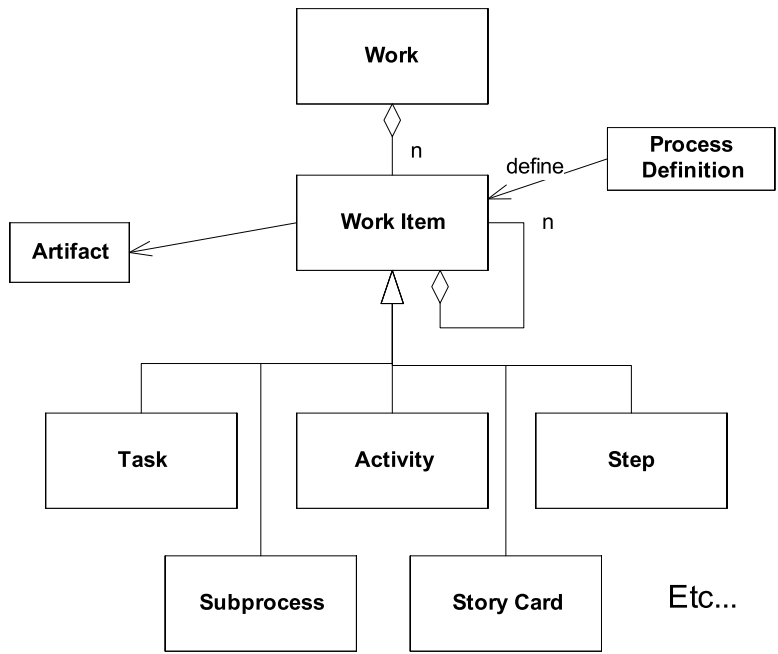

*Figure 7: Work Items come in many different forms and size*

Many work items 'operate' on some concrete, explicit artifact, that is, they use artifacts as input and create or update artifacts. These artifacts, their templates, and the details of the work are part of the project's process, and these templates can evolve over time during a project.

**Product, revisited**
One could claim that the delivered product is in the end the only thing that should matter to the software project manager. What constitutes the product will vary greatly across domains, from tiny software embedded in some device, to large distributed systems, upgraded dynamically weekly, from shrink-wrapped software sold over the internet to "software as a service", from one-off "Kleenex" software rapidly assembled out of a software junkyard to mission-critical applications maintained over 25 years. In the simple cases, the product consists in executable code, often targeted to a specific set of operating systems, accompanied by application data, and some supporting material: user guides, training material, etc. Nowadays products are often expected to run on multiple OS platforms, and several versions of these, as well as supporting different locales: languages and work habits specific to the countries of the user.

A notable concept associated to the product is that of a *release*, which is a product made available to certain parties at given points in time during the project.

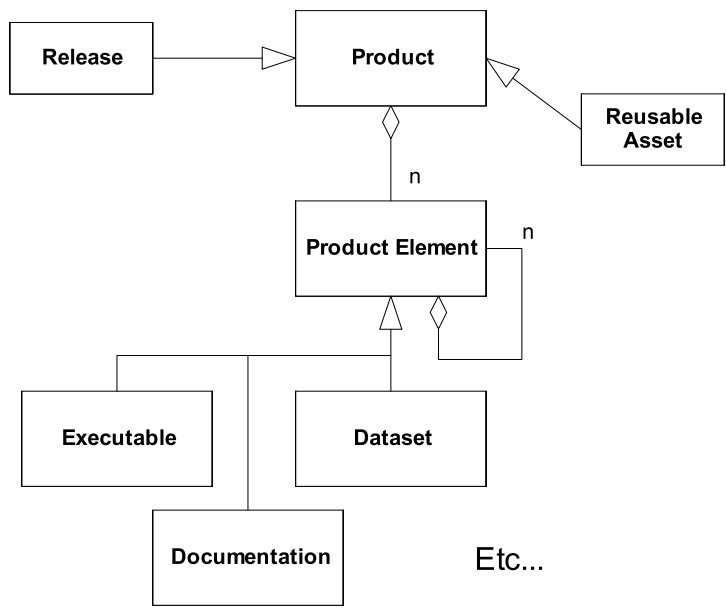

*Figure 8: Elements of a Product vary across types of projects;
a Release and Reusable assets are distinguished forms of Product*

It is around the concept of product that we can discuss issues such as software as an asset, intellectual property rights, and the reuse of software assets from project to project, whether this software is open-source, commercial-off-the-shelf, or proprietary.

**People, revisited**
We've come to realize, through some pains, that software engineering is not primarily a technical issue, but a people issue. Most of the real difficulties in software development, most of the errors and shortcomings are not related to technologies but to the people developing it, their competence, experience and availability, the communication and coordination between these people or teams of people. Therefore the staff component, which is very often ignored in process standards and methods, or abstracted as some kind of vague and perfect agent, plays a major role in our conceptual model. There are several aspects crucial to software project management: the persons themselves, i.e., the individuals, with their knowledge and competence, the roles they play in the software development process: analyst, developer, tester; their organizations in teams, and the allocation of work to people or teams. Notions like *trust* (Moe and Šmite, 2008) and *motivations* (Pink, 2011) play an enormous part and are not captured by the standards cited above.

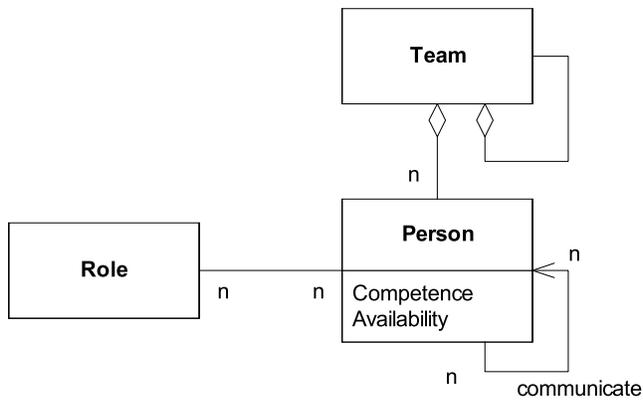

*Figure 9: Persons, teams and roles*

In a large organization, one can envisage team to have roles. It's also around our concept of people that we can discuss issues such as ethics and professional practice.